# A systematic review: Deep learning-based methods for pneumonia region detection

**Xinmei Xu**

Shenzhen College of International Education, Shenzhen, 518043, China

suzy.xu2024@outlook.com

**Abstract.** Pneumonia disease is one of the leading causes of death among children and adults worldwide. In the last ten years, computer-aided pneumonia detection methods have been developed to improve the efficiency and accuracy of the diagnosis process. Among those methods, the effects of deep learning approaches surpassed that of other traditional machine learning methods. This review paper searched and examined existing mainstream deep-learning approaches in the detection of pneumonia regions. This paper focuses on key aspects of the collected research, including their datasets, data processing techniques, general workflow, outcomes, advantages, and limitations. This paper also discusses current challenges in the field and proposes future work that can be done to enhance research procedures and the overall performance of deep learning models in detecting, classifying, and localizing infected regions. This review aims to offer an insightful summary and analysis of current research, facilitating the development of deep learning approaches in addressing treatable diseases.

## 1. Introduction

Pneumonia is the largest cause of death in children around the globe; it accounts for 22% of deaths in children from 1 to 5 years old worldwide, as reported by the Worldwide Health Organization (WHO) in 2019 [1]. Pneumonia is an infection in the lung by bacteria, viruses, or fungi which may inflame the air sacs in the lung with fluids or pus, causing cough and breathing difficulties. Pneumonia could be prevented if diagnosed in time and precisely; otherwise, the unchecked development of pneumonia could be life-threatening. Precisely speaking, diagnosis of pneumonia could face several difficulties. Highly trained and experienced specialists are required to examine the chest X-Ray radiograph (CXR) or computed tomography (CT) image of the scanned lung. Moreover, the examination process could be tedious and may evoke disagreements among radiologists. The positioning of the patient and his or her depth of breath may also affect the appearance of CXR [2]. Due to those difficulties in diagnosis and the dramatic killing power of pneumonia, a rising demand for more effective pneumonia diagnosis methods had promoted the development of more accurate and efficient computer-aided approaches.

In the last decade, computer-aided methods, such as image processing and machine learning approaches, have been explored to improve the accuracy and efficiency of pneumonia diagnosis. Image processing approaches may involve cropping the medical image, extracting lung region, and thresholding the opacity area (for example, from the healthy part of the lung to calculate the ratio of healthy lung area and total lung area) to reach a result [3]. However, machine learning methods, which learn features in images for pneumonia detection, are more efficient and accurate than conventional image processing methods. Among machine learning approaches, deep learning methods, which have more hidden layers, were proven to output more accurate results than machine learning conventions [4]. Mainstream deep learning methods applied to pneumonia detection include one-stage methods, two-stage methods, transfer learning, and ensemble. They have relative advantages and disadvantages in terms of speed, accuracy, and model complexity. Some of the studies yielded high accuracy in the classification and localization of abnormalities. This review discusses deep-learning methods for pneumonia region detection and proposes directions for further research.

**2. Datasets**

Datasets play a crucial role in training accurate models. In the field of pneumonia detection using deep learning-based methods, the diversity, accuracy, sufficiency, and proportionality of datasets, such as chest X-rays and CT scans, are vital in producing and optimizing accurate models. There are currently multiple representative datasets which are readily and publicly available and were downloaded thousands of times to train and optimize deep learning models. Types of these datasets and their detailed information are illustrated in the following section.

*2.1. Types of datasets*
Chest radiography, or CXR, is the most common imaging modality used in clinical pneumonia diagnosis [5]. The X-ray exam allows doctors to see white spots, or infiltrates, in the lungs from the front view to identify an infection. CT scan is another common imaging modality which provides finer details of the lungs from the upper view. Other imaging methods in pneumonia detection include ultrasound, OCT (Optical Coherence Tomography), and MRI (Magnetic Resonance Imaging).

*2.2. Chest X-ray images*
A chest X-ray produces a black-and-white image that presents the organs in the patient's chest. Structures that cannot be penetrated by radiation, mainly bones, appear white; structures that can be penetrated by radiation appear black. An opaque area may indicate the presence of pneumonia. Some chest X-ray datasets of great volumes are publicly available online, presented on platforms such as Kaggle where described data may be provided independently or with its corresponding challenge and can be downloaded. Chest X-ray images were usually collected from healthcare institutes and labelled with one or more lung diseases by experts. Some more detailed datasets may even include bounding boxes which indicate opacity area.

One of the largest publicly available chest x-ray datasets would be the ChestX-ray14 dataset, uploaded by the National Institutes of Health (NIH) on the Kaggle platform (examples are shown in **Figure 1.** Examples of fourteen types of thoracic diseases in the *ChestX-ray14* dataset). The ChestX-ray14 dataset contains 112,120 frontal-view x-ray images collected from 30,805 unique patients [6][7]. In this dataset, images are labelled among 14 thoracic diseases, such as atelectasis, infiltration, fibrosis, and pneumonia, according to the radiological reports of the corresponding images. This dataset was utilized by many researchers. For example, the CheXNet algorithm, which is one of the best-performed algorithms in detecting pneumonia, utilizes the ChestX-ray14 dataset to train, validate, and test the model which detects pneumonia and other thoracic diseases [8].

Another large publicly available chest X-ray dataset is provided with the RSNA Pneumonia Detection Challenge hosted on the Kaggle platform, where the Radiological Society of North America (RSNA) collaborated with NIH to develop this dataset for the challenge [9]. 26,684 unique CXR images labelled with around 37,000 unique patient IDs are divided into 31% with opacity, 41% with no opacity (normal), and 29% other (not normal and no opacity). Bounding boxes of abnormal areas are included in images labelled with pneumonia positive. Patients' further details such as age and sex are also included in the dataset. A study of deep learning for automatic pneumonia detection utilizes this dataset and claims to have demonstrated one of the best results in the RSNA challenge [10].

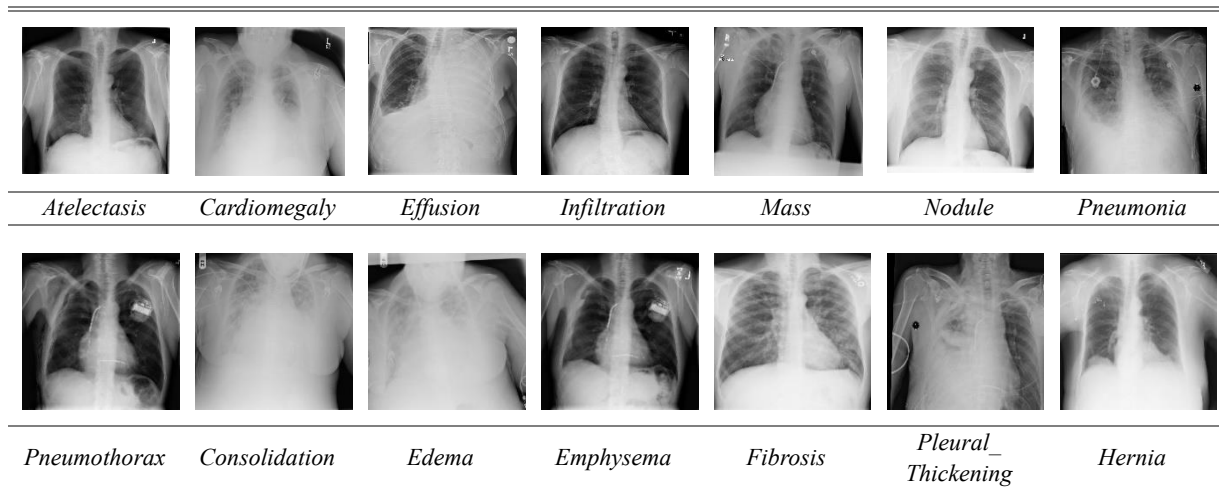

**Figure 1.** Examples of fourteen types of thoracic diseases in the *ChestX-ray14* dataset.

The "Chest X-Ray Images (Pneumonia)" published on the Kaggle platform is another popular dataset which has a high download number of more than 200,000 times. The X-ray images were selected from one to five-year-old paediatric patients from Guangzhou Women and Children's Medical Centre. This dataset contains three folders (train, test, and validation) where images are further divided into two types (normal and pneumonia) in each folder. Its 5,863 X-ray images in total were graded with pneumonia or normal by two expert physicians and the grading was then examined by a third expert to reduce labelling errors [11].

*2.3. CT images*

A CT scan is usually implemented when doctors cannot make a conclusive diagnosis from the patient's CXR results, or when the patient's pneumonia symptoms are considered to be relatively serious. While CXR offers a more general view of the effect of infection on the lung, a CT scan detects more details about the status of the lung, which may include whether other tissues are affected by infection, the severity of infection, and the type of pathogen(s) that infected the lung. Therefore, a CT scanning image dataset usually appears with and correlates to a specific type of pneumonia, such as COVID-19 pneumonia, rather than general pneumonia, leading to the low availability of abundant CT scan datasets. Thus, in studies exploring deep learning methods in pneumonia detection relying upon CT images as datasets, CT scan images may usually be collected from several different healthcare institutes to train deep learning models.

For instance, a widely referenced deep learning study based on CT imaging for COVID-19 pneumonia detection [12] collected 1,396 CT imaging datasets from three different hospitals in which two datasets are publicly available. The first public dataset is described in the form of a research paper [13] and presented on the Kaggle platform [14]. This dataset contains 812 images in total, in which 349 images are COVID-19 CT from 216 patients and 463 images are non-COVID-19 CT. The second public dataset is available in the form of the website, which includes 100 CT scans with ground truth lesion segmentation labelled by physicians from different hospitals [15].

## 3. Object detection workflow

Raw datasets are usually preprocessed before being used to train models. In studies of pneumonia detection using deep learning methods, general data preprocessing techniques include labelling, resizing, augmenting, normalizing, splitting the data, as well as transfer learning.

In the training of the CheXNet model, for example, the CXR images were labelled as either positive (pneumonia present) or negative (all other images with pneumonia absent). The images were downscaled to 224×224 pixels and normalized according to the mean and standard deviation of images in the ImageNet training set. To increase dataset volume, the team also augmented the training data with random horizontal flipping. To evaluate the model, the whole dataset was split into training, validation, and testing without patient overlap between the sets [8].

In another study, the obtained dataset ChestX-ray14 has already been preprocessed by downscaling and labelling [7]. The team further preprocessed the training data by splitting the dataset into training, tuning, and validation without patient overlaps among the partitions [16].

*3.1. One-stage method*
The one-stage method is a type of object detection architecture used in computer vision tasks. One-stage method directly predicts object classes and the coordinates of bounding boxes in a single pass through the network. These methods usually use a convolutional neural network (CNN) to extract features from the whole image and apply a set of anchor boxes to predict object locations and class probabilities. Examples include Single Shot Detector, You Only Look Once (YOLO), RetinaNet, and EfficientDet. One-stage methods are designed to be faster and have less model complexity compared to two-stage methods. However, one-stage methods also yield poorer accuracy compared to two-stage methods.

In one study of pneumonia detection using deep learning methods, several one-stage methods were applied and compared, including RetinaNet and YOLOv3, and the former produces accuracy comparable to that of a two-stage detector by using focal loss [10].

*3.2. Two-stage method*
The two-Stage method, as its name suggests, involves two stages for object detection. In the first stage, these methods generate region proposals, which are potential locations of objects in the image. Techniques such as region proposal networks (RPNs) or selective search can be used to obtain these proposals. In the second stage, the selected region proposals are classified and refined to accurately locate the object. A CNN is typically used to extract features from the entire image in the first stage; in the second stage, a smaller region-based CNN is applied to each proposed region. Examples of two-stage methods include Faster Region Convolutional Neural Networks (Faster R-CNN), and Region-based Fully Convolutional Networks (R-FCN). Two-stage methods usually produce more accurate results with higher model complexity and time consumption compared to one-stage methods.

For instance, the CheXNeXt model, which yielded good results in classifying 14 thoracic diseases, applies a two-stage method. In the first stage, several networks trained on the training set predicted the probability of each of the 14 pathologies in each image. In the second stage, a subset of those networks made up of an ensemble that yielded predictions through calculating the mean of the predictions of every single network [16]. Studies that also applied two-stage methods include the ones using Mask-RCNN and RetinaNet [18][19].

*3.3. Other deep learning methods*
Apart from one-stage and two-stage methods, other typical methods used in the detection of pneumonia include ensemble, transfer learning, CNN, and traditional machine learning. A summary of mainstream deep learning methods in pneumonia detection is presented in **Table 1.** Summary of deep learning methods for pneumonia region detection. CXR images are chest X-ray images, CT images are computed tomography images, which shows the corresponding paper, the dataset used, the type of images, model architecture, results, advantages, and disadvantages.

**Table 1.** Summary of deep learning methods for pneumonia region detection. CXR images are chest X-ray images, CT images are computed tomography images, and OCT images are optical coherence tomography images.

| No. | Papers | Dataset | Type of images | Architecture | Results | Advantages | Limitations |
| --- | --- | --- | --- | --- | --- | --- | --- |
| 1 | [8] | Chest X-ray14 [7] | CXR | 121-layer CNN, transfer learning | Accuracy: 0.7680 | Multiclass classification, localization via heatmap visualization | Lack of incorporation of datasets with a lateral view and with patient history |
| 2 | [16] | Chest X-ray14 [7] | CXR | 121-layer DenseNet | Accuracy: 0.828 The area under the curve (AUC): 0.851 | Multiclass classification, interpretation of network prediction via heatmap, outperformed radiologists on acute diagnoses | Lack of incorporation of datasets with a lateral view and with patient history, low resolution of images in model training, lack of algorithm generalizability due to a single source of the dataset |
| 3 | [20] | Chest X-ray14 [7] LIDC-IDRI [21] | CXR, CT | CXR model: AlexNet, VGG16, VGG19, ResNet50, MAN-SoftMax, MAN-SVM CT model: AlexNet, VGG16, VGG19, ResNet50, MAN-SVM, MAN-KNN, MAN-RF | CXR(MAN-SVM): Accuracy: 0.9680 Sensitivity:0.9697 Specificity:0.9663 CT (MAN-SVM): Accuracy: 0.9727 Sensitivity:0.9809 Specificity:0.9563 | Utilization of both CXR and CT images, the performance of multiple models compared, high accuracy, sensitivity, and specificity | The diversity of datasets can be enhanced by incorporating datasets from other institutions |
| 4 | [22] | Chest X-ray14 [7] | CXR | 169-layer DenseNet and SVM | AUC: 0.8002 | Combination of DenseNet as feature-extractors and SVM as classifier which meliorates the model performance | Lack of infection localization |
| 5 | [23] | Custom [24] | OCT, CXR | Pretrained Inception V3 on ImageNet, adapted for transfer learning | Accuracy: 0.928 Sensitivity: 0.932 Specificity: 0.901 | Localization of infection area using heatmap, model's performance comparable to human experts | Universalization of the algorithm can be improved by using datasets from diverse institutes |
| 6 | [18] | RSNA [9] | CXR | Mask-RCNN | / | Use of thresholds in the background while training | Lack of measures to address dataset imbalance |
| 7 | [25] | Pediatric [11] | CXR | Pretrained MobileNetV2, Xception, and ResNet adapted for transfer learning | Accuracy: 0.764 | Reduction of computation complexity | Classification accuracy can be improved |

**Table 1.** (continued).

| No. | Papers | Dataset | Type of images | Architecture | Results | Advantages | Limitations |
| --- | --- | --- | --- | --- | --- | --- | --- |
| 8 | [17] | RSNA [9] | CXR | CheXNet with NIH pretrain, RetinaNet, YOLOv3, CheXNet+YOLOv3 | F1 Score: 0.40, N/A, 0.81, 0.8385 | Localization of infected region using heatmap, error analysis | F1 Scores can be improved, image quality can be standardized |
| 9 | [10] | RSNA [9] | CXR | SSD RetinaNet with SE-ResNext101 encoder pre-trained on ImageNet | mAP (mean average precision): 0.24781 | Heavy augmentations, multi-task learning with global classification output, postprocessing | Lack of localization of the infected region |
| 10 | [19] | RSNA [9] | CXR | Ensemble of RetinaNet and Mask R-CNN | Recall: 0.793 F1 score: 0.775 | Localization of infected region using bounding boxes | Patient history, and lateral x-ray views can be included to produce a comprehensive diagnosis |

## 4. Discussion

This review paper examines recent pneumonia detection methods based on deep learning, among which there are some challenges yet to be addressed and limitations to be solved. One significant aspect would be datasets. The collection of datasets can be challenging in reality in several steps, and addressing the challenges can largely help to improve the training effects of models. First, the availability of datasets with pneumonia cases is limited. These datasets are usually not readily available or attainable; rather, researchers have to collaborate with medical institutions or hospitals that hold the necessary data. Secondly, since scans are subject to privacy regulations and sensitivity, it can be time-consuming and tangled to obtain the required consent, and permissions, and ensure compliance with data protection laws. Moreover, only trained radiologists are capable of labelling and classifying scans (into normal/pneumonia/other symptoms), which the process can be prone to human error, time-consuming, and labour-intensive due to a lack of qualified professionals. Lastly, the quality and standardization of scans can be difficult to maintain. Scans can vary regarding image quality, presence of artefacts, positioning of patients, and capturing techniques used, posing barriers for models to learn discriminative features effectively. Additionally, the same procedure of exploiting datasets can lead to more obstacles for small-scale research institutions as it requires considerable resources, such as financial investment, computational infrastructures, and a skilled workforce, to build a robust dataset. Hence, collaborations among institutions and related parties are essential and should be encouraged in constructing qualified and robust datasets.

Data preprocessing techniques also affect the performance of the trained model. For instance, the higher the resolution of images, the greater results the model will yield; however, high resolution can also lead to overfitting, greater model complexity, and larger consumption of computational power and time when running the model. There are always trade-offs between available resources and model complexity; thus, it is crucial to explore the right level of complexity that matches the dataset and computational limitations to obtain optimal performance.

In addition, the use of the one-stage method versus the two-stage method impacts accuracy, speed, and model complexity. The one-stage method offers the advantage of faster inference times and thus can be computationally efficient and suitable for real-time applications in diagnosis scenarios. Meanwhile, one-stage methods have limited accuracy in localizing small or overlapping lesions and the capability to handle complex spatial relationships, which can be vital in pneumonia detection. In contrast, two-stage methods often provide better localization accuracy due to their refinement process; this can be beneficial for pneumonia detection tasks requiring precise localization of infected regions and abnormalities. However, two-stage methods are typically more computationally demanding and slower compared to one-stage methods. Thereby, the choice between one-stage or two-stage methods depends on the goals of diagnosis and the technical constraints of computational devices.

## 5. Conclusion

In conclusion, this review paper analysed and examined a wide range of literature on pneumonia detection based on deep learning methods. Through a comprehensive exploration of the existing studies, multiple key findings and contributions have been shown. First, it has become evident that two-stage methods often yield more accurate results than one-stage methods. Second, transfer learning is a mainstream method applied in constructing models to detect pneumonia regions. Additionally, this review has highlighted the importance of robust datasets, choice of data preprocessing techniques and object detection workflow in producing models with high performance.

Furthermore, this review has identified gaps and limitations in the current body of knowledge, suggesting directions for future research. For instance, future research can focus on the integration of several imaging modalities such as chest X-ray and CT scans, along with clinical data such as patient history, which can potentially improve the accuracy of pneumonia detection. Also, the interpretability of the model can be enhanced by exploring techniques to locate and highlight the region of abnormalities on the image to assist doctors in further diagnosis and precise localization. In addition, to enable practical implementation of pneumonia detection models, future research should consider the challenges associated with deploying deep learning systems in clinical scenarios, including addressing issues of data privacy, regulatory compliance, integration with contemporary healthcare infrastructures, and validation in real-world settings.

Overall, this review paper has contributed to the existing body of literature by providing an in-depth analysis of pneumonia detection based on deep learning methods. By synthesizing the findings from various research, it also proposes future research directions on improving the accuracy, efficiency, interpretability, and clinical utility of deep learning approaches. It is hoped that this review will serve as a valuable resource for researchers and practitioners, promoting future investigation and boosting the development of innovative approaches to leverage existing findings to ultimately facilitate clinical outcomes of the treatable disease pneumonia.